\documentclass[twocolumn,amsmath,amssymb,floatfix,nofootinbib,superscriptaddress]{revtex4}
\usepackage{graphicx,graphics,times,color}
\usepackage{bm}
\usepackage{longtable}
\usepackage{color}
\usepackage{lipsum}
\usepackage{epstopdf}
\usepackage{amsmath}
\usepackage{appendix}

\def\be{\begin{equation}}
\def\ee{\end{equation}}
\def\ba{\begin{eqnarray}}
\def\ea{\end{eqnarray}}
\def\la{\langle}
\def\ra{\rangle}

\begin{document}

\title{ Measurement-Assisted Quantum Communication in Spin Channels with Dephasing }

\author{Abolfazl Bayat}
\affiliation{Department of Physics and Astronomy, University College London, 
United Kingdom}

\author{Yasser Omar}
\affiliation{Physics of Information Group, Instituto de Telecomunica\c{c}\~oes, Lisbon, Portugal}
\affiliation{CEMAPRE, ISEG, Universidade de Lisboa, Portugal}

\date{\today}

\begin{abstract}
We propose a protocol for countering the effects of dephasing
in quantum state transfer over a noisy spin channel weakly coupled to the
sender and receiver qubits. Our protocol, based on performing regular
global measurements on the channel, significantly suppresses the nocuous
environmental effects and offers much higher fidelities than the
traditional no-measurement approach. Our proposal can also operate as a
robust two-qubit entangling gate over distant spins. Our scheme counters
any source of dephasing, including those for which the well established
dynamical decoupling approach fails. Our protocol is probabilistic,
given the intrinsic randomness in quantum measurements,
but its success probability can be maximized by adequately
tuning the rate of the
measurements.
\end{abstract}


\maketitle

\section{Introduction}

Over the last decade there have been several proposals for exploiting the natural time evolution of many-body systems for short-range quantum communication between separated registers of a quantum network \cite{bose-review,bayat-review-book}.  Only recently the first experimental realizations of quantum state transfer through time evolution of many-body systems have been achieved in NMR \cite{state-transfer-NMR}, coupled optical fibers  \cite{kwek-perfect-transfer} and cold atoms in optical lattices \cite{Bloch-spin-wave,Bloch-magnon}. One of the major challenges in the realization of all quantum processes is dephasing, which destroys the coherent superpositions of states and results in classical mixtures \cite{Breuer-OpenQS-2002}. The origin of dephasing is the \emph{random} energy fluctuations induced on qubit levels by random magnetic and electric fields in the environment. Dynamical decoupling \cite{Viola-Lloyd-Knill-DD}, as an open-loop control technique, has been developed to overcome dephasing through performing regular instantaneous control pulses \cite{Burgarth-DD-2014} and has been very effective in designing long-time memory cells \cite{Khodjasteh-DD-2013} and quantum gates \cite{Lidar-gate-DD-2010}.  Nevertheless, the dynamical decoupling technique is only effective for static (or very slow time-varying) random fields, such as the hyperfine interaction in solid state quantum dot qubits \cite{Yacoby-DD-2011}. In particular, when the fluctuations in qubit levels are time dependent or in the case of Markovian decoherence, explained by a master equation, dynamical decoupling fails to compensate decoherence effects in the system \cite{Burgarth-decoupling-2014}.

In a simple quantum state transfer scenario with a uniform spin chain, the evolution is dispersive and thus the quality of transport decreases by increasing the size  \cite{bose-review}. Hence, to realize perfect state transfer, spin chains with engineered couplings were proposed \cite{christandl}, and some modifications may also allow them to operate independently of their initialization \cite{DeFranco-perfect} (see Ref.~\cite{Kay-review} for a detailed review on perfect state transfer). One may also get arbitrary perfect state transfer in uniform chains using dual-rail systems \cite{burgarth-dual-rail}, $d$-level chains \cite{bayat-dLevel-2014} or arrays of prime number of qubits \cite{Severini, Omar-Prime}. In free fermionic systems, one gets arbitrarily high fidelities by engineering the two boundary couplings \cite{leonardo}. Alternatively, one may use intermediate spins as interaction mediators between a sender and a receiver which are off-resonant from the channel by either using weak couplings \cite{weak-coupling} or strong local magnetic fields \cite{Perturbation-magnetic}. In these scenarios the intermediate spins are only virtually populated \cite{weak-coupling,Perturbation-magnetic} and the dynamics is governed by an effective Hamiltonian between the two ending spins which offers perfect state transfer in the absence of dephasing.

\begin{figure} \centering
    \includegraphics[width=8cm,height=4.5cm,angle=0]{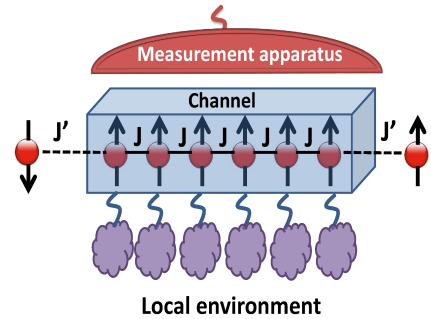}
    \caption{A spin channel with uniform couplings $J$, under the effect of local dephasing, is weakly coupled to the sender and receiver qubits. Regular collective measurements on the channel counters the effect of dephasing resulting in high transmission fidelities. }
     \label{fig1}
\end{figure}

Projective measurements are essential elements of quantum technologies, for example for teleportation \cite{Bennet-teleportation}, measurement-based quantum computation \cite{Briegel-MBQC}, entangling macroscopic atomic ensembles \cite{Sherson-measurement},  many-body state engineering \cite{Sherson-state_engineering} or entanglement generation between superconducting qubits within a meter of distance \cite{Seddiqi-SC-measurement}.
In particular, for cold atoms in optical lattices, nondemolition quantum measurements have been proposed for the creation and the detection of spin-spin correlations \cite{Lewenstein-measurement}. In such systems, quantum measurements can also be combined with the natural time evolution to engineer complex quantum states \cite{Sherson-state_engineering}, as well as quantum communication in spin chains \cite{bayat-song2011,bayat-pouyandeh2014}.
Furthermore, continuous measurements leading to quantum Zeno effect \cite{Sudarshan-zeno-effect-1977, Pascazio-Zeno-2008} can be used to totally or partially freeze the evolution of the system and even suppress decoherence, namely by restricting the coherent evolution to a reduced Hilbert space \cite{Smerzi-Zeno-NatureComm,Pleastina-Zeno} or by preventing unstable states from decaying \cite{Kondo-Zeno-2014}.

In this paper, we show how global measurements, performed regularly on the spin channel, can counter the effect of dephasing, even in scenarios where dynamical decoupling fails, offering high transmission fidelities. Our mechanism uses measurement as a mean for purification of the system, and thus countering the effect of dephasing, without entering the Zeno zone in which the dynamics is frozen.

\section{Model}

Let us consider a uniform spin chain (our channel) in which spins are labelled from $2$ to $N-1$, where $N$ is even,  with the interacting Hamiltonian of the following form
\begin{equation} \label{Hch}
  H_{ch} = J \sum_{k=2}^{N-2} \{ \sigma_k^+ \sigma_{k+1}^- + \sigma_k^- \sigma_{k+1}^+ \}
\end{equation}
where $J$ is the exchange coupling and $\sigma_k^+$ and $\sigma_k^-$ are the Pauli spin ladder operators acting on site $k$.

The channel is initialized in the ferromagnetic state $|\mathbf{0}_{ch}\ra=|0,0,...,0\ra$, in which all spins are aligned. Two extra spins, i.e.\ qubits $1$ and $N$, are located at both ends of the channel. At $t=0$ these two qubits are suddenly coupled to the ends of the channel, as shown in Fig.~\ref{fig1}, via the following Hamiltonian
\begin{equation} \label{H_I}
  H_I = J' (\sigma_1^+ \sigma_{2}^- + \sigma_1^- \sigma_{2}^+ + \sigma_{N-1}^+ \sigma_{N}^-+ \sigma_{N-1}^- \sigma_{N}^+)
\end{equation}
where $J'$ is the boundary spin couplings to the channel, and throughout this paper it is assumed to be much smaller then the spin couplings in the channel, namely $J'\ll J$. The total Hamiltonian of the system is thus $H=H_{ch}+H_I$. Qubit $1$ encodes the state to be sent, and is initialized in an arbitrary (possibly unknown) state $|\psi_s\ra=\cos(\theta/2)|0\ra +e^{i\phi}\sin(\theta/2)|1\ra$, and qubit $N$, the receiver spin, is initialized in the state $|0\ra$. So, the initial state of the whole system can be written as
\begin{equation}
\rho(0) = |\psi_s\ra \la \psi_s| \otimes  |\mathbf{0}_{ch}\ra \la \mathbf{0}_{ch}| \otimes |0\ra \la 0|.
\end{equation}

Generally the channel is not well isolated from the environment and might be disturbed by the effect of surrounding fluctuating magnetic or electric fields, which induce random level fluctuations in the system which then result in dephasing. For fast and weak random field fluctuations, one can get a master equation \cite{Breuer-OpenQS-2002} for the evolution of the system  as
\begin{equation}\label{super_operator_t}
\dot{\rho}(t)=-i[H,\rho(t)]+\gamma \sum_{k=2}^{N-2}  \{ \sigma_k^z \rho(t) \sigma_k^z -\rho(t) \}
\end{equation}
where the first term in the right hand side is the unitary Schr\"{o}dinger evolution and the second term is the dephasing which acts on the channel qubits with the rate $\gamma$. To see the quality of the quantum state transfer, one may compute the density matrix of the last site by tracing out the other spins
\begin{equation}\label{rho_N}
  \rho_N(t) = Tr_{\widehat{N}} \rho(t)= \xi_{t} [|\psi_s\ra \la \psi_s|],
\end{equation}
where $Tr_{\widehat{N}}$ means tracing over all spins except qubit $N$, and $\xi_{t}$ is the super-operator determining the linear relationship between the input and the output of the channel.
Then one can compute the fidelity of the output as $F(\theta,\phi;t)=\la \psi_s|\rho_N(t)|\psi_s\ra$.
In order to have an input-independent parameter, one can compute the average fidelity with respect to all possible input states on the surface of the Bloch sphere as
\begin{equation} \label{F_av1}
  F^{av}(t)= \int F(\theta,\phi;t) d\Omega,
\end{equation}
where $d\Omega$ is the normalized $SU(2)$ Haar measure. A straightforward calculation shows that in the master equation (\ref{super_operator_t}) in which the total magnetization is conserved, the average fidelity can be written as \cite{bayat-xxz}
\begin{equation} \label{F_av2}
  F^{av}(t)= \frac{1}{2}+\frac{1}{6}F^{exc}(t)+\frac{1}{3}F^{coh}(t).
\end{equation}
where
\begin{eqnarray} \label{F_av3}
  F^{exc}(t)&=& \la 1 | \xi_{t} [|1\ra\la 1|] |1\ra, \cr
  F^{coh}(t)&=& | \la 0 | \xi_{t} [|0\ra\la 1|] |1\ra |.
\end{eqnarray}
While $F^{exc}(t)$ quantifies how well this channel can transmit classical excitations, the parameter $F^{coh}(t)$ accounts for the quantum coherence preservation of the channel. In particular we are interested in a special time $t=t_m$ at which the average fidelity peaks for the first time $F^{av}_m=F^{av}(t_m)$.  \\

\begin{figure} \centering
    \includegraphics[width=9cm,height=8cm,angle=0]{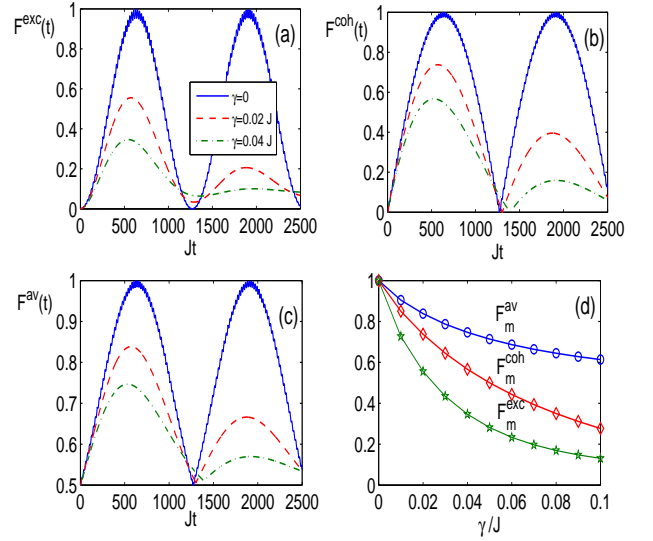}
    \caption{  Different fidelities versus time for various dephasing rates $\gamma$ in a chain of length $N=12$ with $J'=0.05J$: (a) $F^{exc}(t)$; (b) $F^{coh}(t)$; (c) $F^{av}(t)$. (d) The maximum fidelities as functions of the dephasing rate $\gamma$ in a chain of length $N=12$.}
     \label{fig2}
\end{figure}

\section{Effective Hamiltonian ($\gamma=0$)}

In the case of no dephasing, where $\gamma=0$, the total energy of the system is conserved and for the specific choice of $J'\ll J$ the channel is only virtually populated during the evolution. Nevertheless, these virtual excitations in the channel mediate an effective Hamiltonian between the  qubits $1$ and $N$, which can be computed using adiabatic elimination
\begin{equation}\label{H_eff}
  H_e = J_{e} (\sigma_1^+ \sigma_N^-+ \sigma_1^- \sigma_N^+)
\end{equation}
where
\begin{equation}\label{J_eff_1}
  J_{e}=(-1)^{N/2} \frac{J'^2}{J}
\end{equation}
is the effective coupling between the qubits $1$ and $N$ mediated through the channel. The  effective Hamiltonian is valid only when the coupling $J'$ is much smaller than the energy gap of the channel which then implies
\begin{equation}\label{J_eff}
  J' \ll \frac{\pi J}{N}.
\end{equation}
Throughout this paper we always work in this regime. Interestingly, for those chains where the coupling $J'$ satisfies this criterion, the effective Hamiltonian is independent of $N$, apart from an irrelevant sign which has no effect for transport properties.
Considering the effective Hamiltonian $H_e$, one can easily show that in the absence of dephasing (i.e.\ $\gamma=0$) the average fidelity in Eq.~(\ref{F_av2}) takes the form of
\begin{equation}\label{Fav_eff}
  F_{av}=\frac{1}{3}+\frac{(1+|\sin(J_et)|)^2}{6}.
\end{equation}
The average fidelity thus reaches its maximum, i.e.\ $F_{av}^m=1$, at the time
\begin{equation}\label{t_m_topt}
  t_m=\frac{\pi}{2J_e}.
\end{equation}

Moreover, if the qubit $N$ is not initialized in $|0\ra$ then the unitary evolution of the system, governed by the free fermionic Hamiltonian $H_I$, performs a two-qubit entangling gate at $t=t_m$, essential for universal quantum computation \cite{Nielsen-Chuang-2000}, between the qubits $1$ and $N$.

It is worth mentioning that, since $J_e$ is much smaller than $J$, the reduction to an effective two-qubit system comes at the price of significantly increasing time operations and decreasing the energy gap (and thus thermal stability). This may lead to a significant experimental challenge, but some systems with large exchange coupling $J$ (like composite systems \cite{Sahling-composite2015}), or those with large coherence times (e.g.\ cold atoms in optical lattices), can be ideal venues for the realization of such effective dynamics. In fact, high order processes in optical lattices, described by an effective Hamiltonian, have already been realized \cite{Bloch-effective-experiment} for counting the atom numbers in a Coulomb-blocked-like scenario. Moreover, a more advanced scheme, based on effective higher order tunnelings, for state preparation, gate operation and particle transfer has been proposed in Ref.~\cite{Sherson-effective-proposal}.

\section{Effect of dephasing ($\gamma \neq 0$)}

In the presence of dephasing, namely nonzero $\gamma$, the energy is no longer conserved and thus the excitations can leak from qubit $1$ to the channel leading to imperfect transfer between the sender and receiver qubits. In Figs.~\ref{fig2}(a)-(c) the fidelities $F^{exc}(t)$, $F^{coh}(t)$ and $F^{av}(t)$ are plotted respectively as functions of time for different values of $\gamma$ in a chain of $N=12$. As the figures clearly show, by increasing the dephasing rate the quality of transmission goes down for all the fidelities. To see how destructive dephasing is, in Fig.~\ref{fig2}(d) we plot the maximum fidelities $F^{exc}_m$, $F^{coh}_m$ and $F^{av}_m$, at $t=t_m$ when their first peak occurs, as functions of the dephasing rate $\gamma$, in a chain of length $N=12$. As the figure shows, the fidelities all decay exponentially with dephasing rate $\gamma$, as expected for the master equation (\ref{super_operator_t}).

\begin{figure} \centering
    \includegraphics[width=9cm,height=8cm,angle=0]{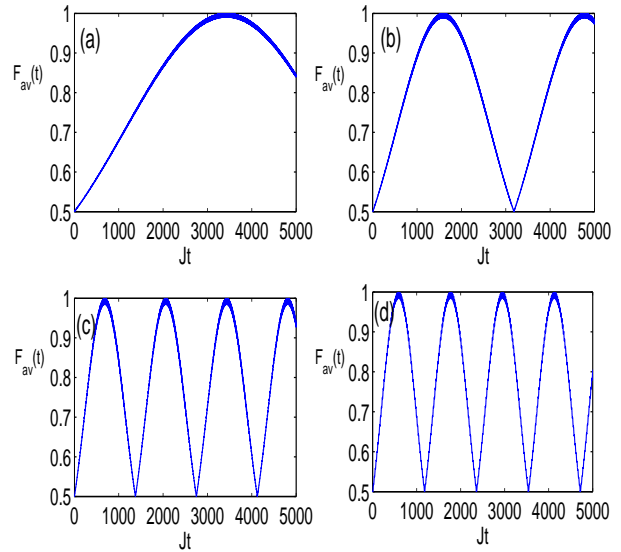}
    \caption{ The average fidelity versus time in a chain of length $N=12$ for zero dephasing (i.e.\ $\gamma=0$). The time intervals between subsequent measurements are: (a) $\tau=6/J$; (b) $\tau=7/J$; (c) $\tau=10/J$; (d) $\tau=20/J$. While the upper panels are inside the Zeno zone, corresponding to slow dynamics, the lower panels are outside that region and thus show faster dynamics.  }
     \label{fig3}
\end{figure}

\section{Regular measurements ($\gamma=0$)}

We now consider a series of global projective measurements on the qubits of the channel at regular time intervals in the absence of dephasing (i.e. $\gamma=0$). The corresponding projection operators are
\begin{eqnarray}\label{Projectors}
M_0&=&|\mathbf{0}_{ch}\ra \la \mathbf{0}_{ch}| \cr
M_1&=& I-M_0,
\end{eqnarray}
where $I$ stands for identity.
If the outcome of the measurement is $M_0$ (i.e.\ the channel is found in $|\mathbf{0}_{ch}\ra$) then the measurement is regarded as successful, otherwise the protocol fails.

To see how the measurements affect the transfer mechanism, we assume that global projective measurements are performed regularly on the channel, at time intervals of $\tau$. Immediately after the $k$'th successful measurement, the state of the system is 
\begin{equation}\label{rho_tau}
  \rho^{(k)}=\frac{M_0\xi_{\tau}[\rho^{(k-1)}] M_0}{Tr\{ M_0\xi_{\tau}[\rho^{(k-1)}] M_0 \}},
\end{equation}
where in this iterative equation $\rho^{(0)}=\rho(0)$, and all $k$ consecutive measurements are assumed to be successful. The probability of successful  measurement at iteration $k$ is
\begin{equation}\label{psuc_k}
  p^{(k)}=Tr\{ M_0\xi_{\tau}[\rho^{(k-1)}] M_0 \}.
\end{equation}
In fact, since the whole protocol fails even if a single measurement yields $M_1$ as its outcome, it makes sense to define the probability of success at the optimal time $t_m$ as the product of all consecutive probabilities until then as
\begin{equation}\label{psuc_k}
  P_{suc}=\Pi_{k=1}^j p^{(k)} {\hskip 0.6 cm} \text{such that: } j\tau \leq t_m < (j+1)\tau.
\end{equation}
After the successful measurement at iteration $k$, the evolution of the system follows the
Eq.~(\ref{super_operator_t}) with the initial state $\rho^{(k)}$.

\begin{figure} \centering
    \includegraphics[width=9cm,height=4.5cm,angle=0]{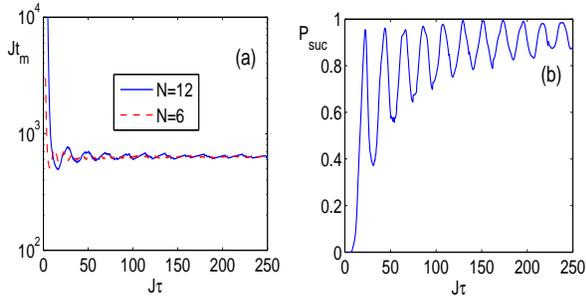}
    \caption{ (a) The optimal time $J t_m$ versus measurement time $J\tau$ for chains of length $N=6$ and $N=12$ without dephasing. (b) The probability of success $P_{suc}$ versus measurement time $J\tau$ in a chain of length $N=12$. In both figures $J'=0.05 J$}
     \label{fig4}
\end{figure}

A very high measurement rate, i.e.\  $\tau/t_m \ll 1$, freezes the dynamics due to the quantum Zeno effect \cite{Sudarshan-zeno-effect-1977}. To see this  effect more clearly, we consider a system of length $N=12$ without dephasing (i.e.\ $\gamma=0$) and plot the average fidelity versus time for different values of $\tau$ in Figs.~\ref{fig3}(a)-(d). As it is evident from these plots, the dynamics is slow for small $\tau$ (high rate of measurement) and it becomes faster by increasing $\tau$ (decreasing the measurement rate).

To have a better insight about the Zeno effect, in Fig.~\ref{fig4}(a) the transfer time $t_m$, at which the fidelity peaks, is depicted as a function of $\tau$. As the figure clearly shows, for small $\tau$ the time $t_m$ is very large, which indicates that the dynamics is practically frozen as predicted by the Zeno effect. By increasing $\tau$, the optimal time $t_m$ decreases and eventually oscillates around $t_m=\pi/(2J_e)$, determined by the effective Hamiltonian (\ref{H_eff}).  The transition between the Zeno and non-Zeno dynamics is determined by two time scales: (i) The time interval between two subsequent measurement, namely $\tau$; (ii) the time scale needed for virtually exciting the channel, namely $\sim 1/J'$, which can only happen with very low probability. As it is evident from Fig.~\ref{fig4}(a), the transition from Zeno to non-Zeno zone happens when $\tau\sim 1/J'$. For smaller values of $\tau$ system enters the Zeno regime and for larger values of $\tau$ the dynamics of the system is not frozen and our protocol can be applied. The number of projective measurements during the transfer time is simply determined by $t_m/\tau$.
In Fig.~\ref{fig4}(b) the success probability $P_{suc}$  is plotted as a function of $\tau$, which is zero in the Zeno regime and rises by increasing $\tau$ and clearly shows resonance behaviour such that, for specific values of $\tau$, the $P_{suc}$ is almost one, meaning that all the consecutive measurements are successful.

Indeed, in the absence of dephasing there is no point in performing regular measurements on the system, as the effective Hamiltonian already achieves perfect state transfer. However, our analysis in this section provides the foundation for the next section in which the operation of  regular measurements compensates the destructive effect of dephasing. Furthermore, the nature of the oscillations in $P_{suc}$ is analysed and discussed in section \ref{sec_Explanation}, where we solve analytically the case of a short chain of length $N = 4$.

\begin{figure} \centering
    \includegraphics[width=9cm,height=4.5cm,angle=0]{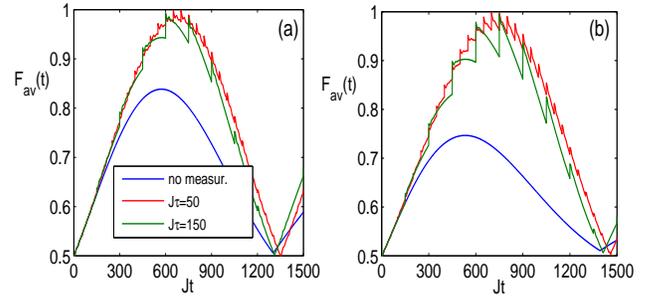}
    \caption{Comparison of the average fidelity (as a function of time) in the no-measurement scenario with the case of our measurement protocol (the latter, for two values of $\tau$), in a chain of length $N=12$. The dephasing rate is: (a) $\gamma=0.02 J$;  and (b) $\gamma=0.04 J$. The improvement in fidelity by performing measurements is very evident in the figures. }
     \label{fig5}
\end{figure}

\section{Tackling dephasing with measurements}

As previously discussed, by increasing the dephasing rate $\gamma$  the average fidelity decays exponentially (see Fig.~\ref{fig2}(d)). Performing regular measurements may improve the fidelity as it purifies the system by projecting the channel into a pure state $|\mathbf{0}_{ch}\ra$. This is indeed the case, as we show in Figs.~\ref{fig5}(a) and (b), where we plot the the average fidelity $F_{av}(t)$ as a function of time for two different values of $\gamma$ respectively. In each of these figures, the dynamics without measurement is compared with the case of our measurement strategy, namely for two different measurement time intervals $\tau$. As the figures clearly show, the attainable fidelity significantly improves by performing measurements. The jumps in the fidelity curves are due to the purification after each successful measurement.

\begin{figure} \centering
    \includegraphics[width=9cm,height=8cm,angle=0]{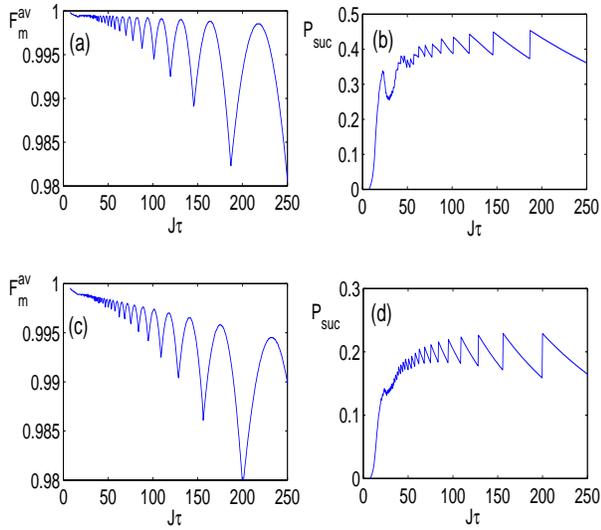}
    \caption{ The maximum average fidelity $F^{av}_m$ and the probability of success $P_{suc}$ as functions of $J\tau$ for a chain of $N=12$ with $J'=0.05 J$ for: (a) and (b) $\gamma=0.02J$; and (c) and (d) $\gamma=0.04J$. In all these figures the transfer time remains almost the same as Eq.~(\ref{t_m_topt}) which will be $t_m \simeq 200 \pi$. }
     \label{fig6}
\end{figure}

In Fig.~\ref{fig6}(a) we show how the maximum average fidelity $F^{av}_m$ varies with the measurement time $\tau$ in a chain of length $N=12$ and $\gamma=0.02$, for which the no-measurement scenario gives $F^{av}_m=0.84$. As the figure shows, by performing regular measurements, the fidelity can go over $0.995$ and shows oscillatory resonance features by varying $\tau$ without going below $0.98$. In Fig.~\ref{fig6}(b) the probability of success $P_{suc}$ is plotted versus $\tau$. As the figure shows, in the Zeno zone (i.e.\ very small $\tau$) $P_{suc}$ is very small, but then rises quickly and eventually fluctuates around an asymptotic value (here around $0.4$). In Figs.~\ref{fig6}(c) and (d) the same quantities are plotted for the same chain but now with $\gamma=0.04$, in which the no-measurement scenario gives $F^{av}_m \simeq 0.75$. As the figure shows $F^{av}_m$ can again be larger than $0.99$, with the price that $P_{suc}$ is going down to $\sim 0.2$. The results evidently show that performing regular measurements improves the fidelity significantly with the price paid for $P_{suc}$. In particular, from Figs.~\ref{fig6}(b) and (d) we can see that a very few ($\sim 4$) measurements are enough to improve the fidelity above $0.99$ in a chain of $N=12$.

\begin{figure} \centering
    \includegraphics[width=9cm,height=4.5cm,angle=0]{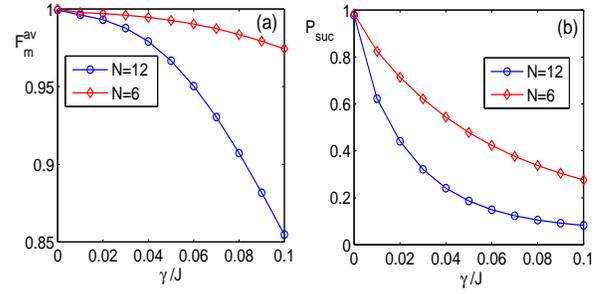}
    \caption{ (a) and (b): The maximum average fidelity $F^{av}_m$ and the probability of success $P_{suc}$ as  functions of $\gamma/J$  for chains of length $N=6$ and $N=12$ with $J'=0.05 J$, and with the measurement time fixed to $J\tau=150$ allowing for four measurements as $Jt_m \simeq 200 \pi$ is determined by Eq.~(\ref{t_m_topt}). }
     \label{fig7}
\end{figure}

In order to see how the average fidelity $F^{av}_m$ scales with the dephasing rate $\gamma$, we fix the length $N$ and the measurement time $\tau$ and plot $F^{av}_m$ as a function of $\gamma$ in Fig.~\ref{fig7}(a). As it is clear from this figure, by choosing $\tau=150/J$ (i.e.\ allowing only four measurements during the evolution), the average fidelity stays very high ($\sim 0.86$) even for very strong dephasing  $\gamma=0.1J$ showing a significant improvement in comparison with the no-measurement scenario, for which $F^{av}_{m}\simeq 0.6$. In Fig.~\ref{fig7}(b) we plot the success probability $P_{suc}$ versus the  dephasing rate $\gamma$, for the same chains and measurement time $\tau$, which shows that increasing dephasing reduces the chance of success, as expected. Naturally, achieving a very high fidelity has a price in terms of $P_{suc}$, but our protocol offers high fidelities for reasonable probabilities of success.

\begin{figure} \centering
    \includegraphics[width=9cm,height=4.5cm,angle=0]{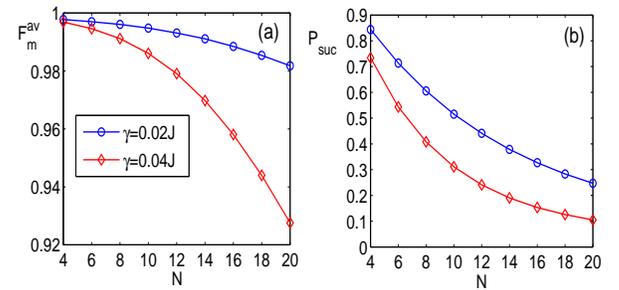}
    \caption{ (a) and (b): The maximum average fidelity $F^{av}_m$ and the probability of success $P_{suc}$ versus $N$ for dephasing rates of $\gamma=0.02J$ and $\gamma=0.04 J$ with $J'=0.05 J$ when the measurement time is fixed to $J\tau=150$ allowing for four measurements as $Jt_m \simeq 200 \pi$ is determined by Eq.~(\ref{t_m_topt}). }
     \label{fig8}
\end{figure}

To finalize our analysis we also study the performance of our protocol for different chain lengths $N$. In Fig.~\ref{fig8}(a) we plot $F^{av}_m$ as a function of $N$ for two values of $\gamma$ when $\tau$ is fixed. This figure shows that the maximum average fidelity decays very slowly by increasing $N$, though, as expected, its decay becomes faster by increasing the dephasing rate $\gamma$. In Fig.~\ref{fig8}(b) the success probability $P_{suc}$ is depicted versus $N$ for the same dephasing parameters, which shows steady decay by increasing length. Similarly to the fidelity, the probability of success also decreases by increasing $\gamma$.

Finally, note that our proposed mechanism is robust against several imperfections, including imprecise measurement timings as well as other types of decoherence. It is worth mentioning that when the effective two-spin Hamiltonian is valid (i.e.\ when the condition in  Eq.~(\ref{J_eff}) is satisfied) the channel is only virtually populated. This means that the proposed protocol works even if the measurement time $\tau$ is not tuned properly or varies from one measurement to another.

Apart from dephasing, the system might be affected by more complex noises which do not conserve the number of excitations. Mathematically this means that the Lindblad operators in the master equation (\ref{super_operator_t}) may include other Pauli operators such as $\sigma^+$, etc. In fact, such operators may induce an excitation in the channel, which then results in failure of the measurement. However, our protocol will still be applicable, but the probability of success $P_{suc}$ will decrease. In order to have a higher probability of success for such noises, one has to use smaller values of $\tau$ (but still not within the Zeno zone).

\begin{figure} \centering
    \includegraphics[width=7cm,height=6.5cm,angle=0]{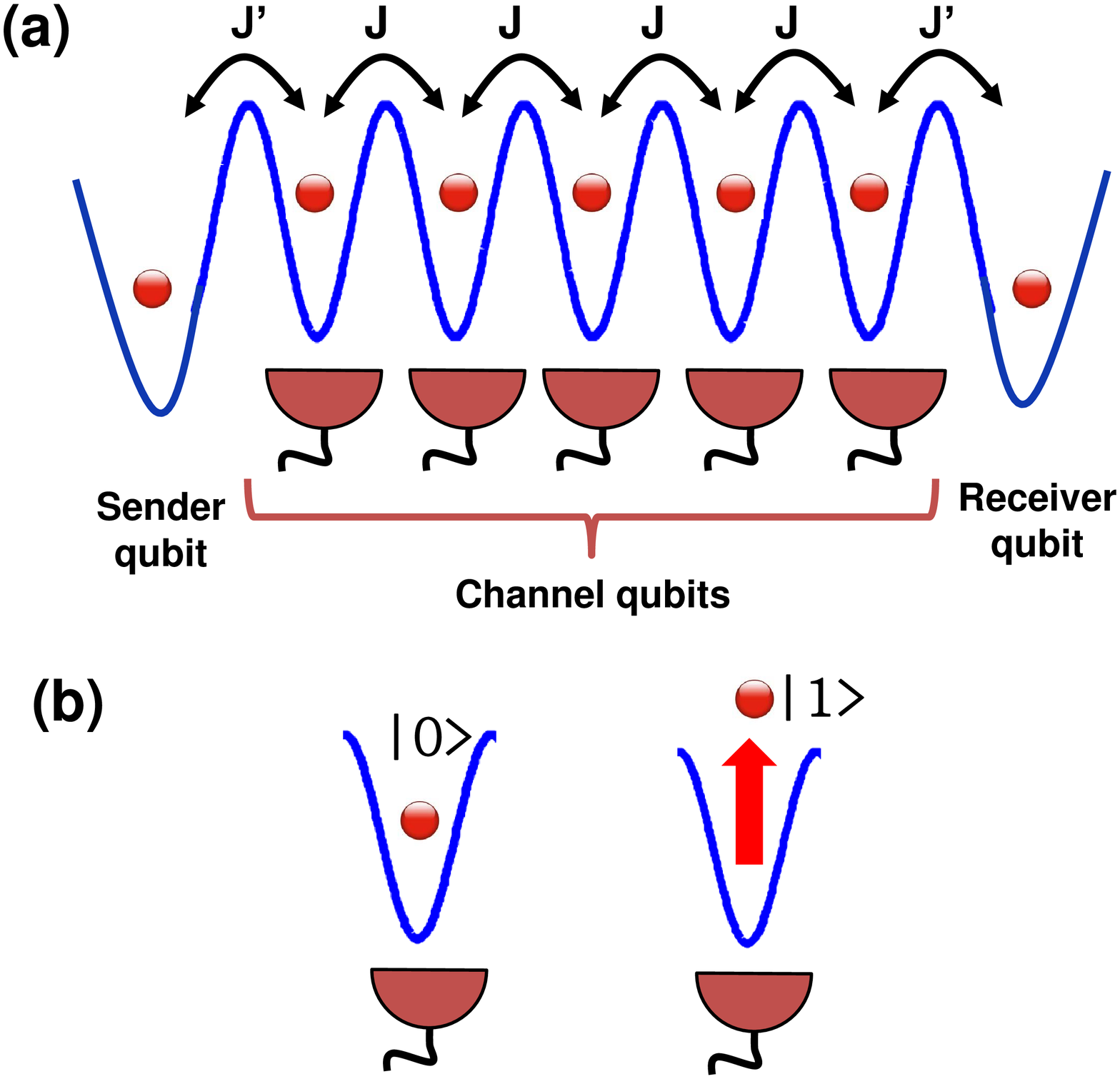}
    \caption{ (a) A possible implementation of our protocol using an optical lattice setup. The particles corresponding to the sender and receiver qubits  feel a deeper potential (and thus a weaker interaction) due to an extra laser beam focused on those sites. (b) For the spin measurements, an intense focused laser beam is coupled to the atomic level $|1\ra$, and the radiation pressure forces the atom out of the lattice if it is in that state. A subsequent fluorescent picture of the lattice will show whether the site is empty or full and the measurement is accomplished. }
     \label{fig9}
\end{figure}

\section{Understanding the resonances in the success probability $P_{suc}$}\label{sec_Explanation}

In order to illuminate the nature of the oscillatory behaviour of the probability of success $P_{suc}$ in time, we consider an isolated (i.e.\ $\gamma=0$) short chain of four spins ($N=4$) which can be solved analytically. In the single excitation subspace, the eigenvalues and eigenvectors of the system are
\begin{eqnarray}\label{Eigs_4spin}
  |E_1\ra &=& \frac{|\mathbf{1}\ra-\alpha |\mathbf{2}\ra + \alpha |\mathbf{3}\ra -|\mathbf{4}\ra}{\sqrt{2(\alpha^2+1)}},  E_1=-\alpha J'      \cr
  |E_2\ra &=& \frac{\alpha |\mathbf{1}\ra- |\mathbf{2}\ra - |\mathbf{3}\ra + \alpha |\mathbf{4}\ra}{\sqrt{2(\alpha^2+1)}}, E_2=J-\alpha J'    \cr
  |E_3\ra &=& \frac{\alpha |\mathbf{1}\ra+ |\mathbf{2}\ra - |\mathbf{3}\ra - \alpha |\mathbf{4}\ra}{\sqrt{2(\alpha^2+1)}}, E_3=-(J-\alpha J') \cr
  |E_4\ra &=& \frac{|\mathbf{1}\ra+\alpha |\mathbf{2}\ra + \alpha |\mathbf{3}\ra +|\mathbf{4}\ra}{\sqrt{2(\alpha^2+1)}},   E_4=\alpha J'
\end{eqnarray}
where $E_k$ and $|E_k\ra$, for $k=1,2,...,4$, are the eigenvalues and eigenvectors of the Hamiltonian respectively, and $|\mathbf{k}\ra$ represents an excitation $|1\ra$ at site $k$ while the other sites are all in $|0\ra$. Moreover, the dimensionless parameter $\alpha$ is defined as
\begin{equation} \label{Alpha_parameter}
  \alpha=\frac{J+\sqrt{J^2+4J'^2}}{2J'}.
\end{equation}
By initializing the whole system in the state $|\Psi(0)\ra=|\mathbf{1}\ra$, one can compute the time evolution of the system in later times as
\begin{equation} \label{Psi_t_4spin}
  |\Psi(t)\ra=e^{-iHt}|\Psi(0)\ra =\sum_{k=1}^4 e^{-iE_kt}|E_k\ra \la E_k|\mathbf{1}\ra.
\end{equation}
The probability of finding the channel (here, sites $2$ and $3$) in the state $|\mathbf{0}_{ch}\ra$ after performing one projective measurement is
\begin{equation} \label{Porb_suc_4spin}
  p^{(1)}=|\la \mathbf{1}|\Psi(t)\ra|^2 + |\la \mathbf{4}|\Psi(t)\ra|^2.
\end{equation}
By inserting the exact form of the eigenvectors from Eq.~(\ref{Eigs_4spin}) in the dynamics of Eq.~(\ref{Psi_t_4spin}) one gets
\begin{equation} \label{Porb_suc_4spin}
  p^{(1)}=\frac{1+\alpha^4+2\alpha^2 \cos[(J-2\alpha J')t]}{(\alpha^2+1)^2}.
\end{equation}
These results are valid for all values of $J'$. However, as previously discussed, we are interested in the limit of $J'\ll J$. In this limit, $\alpha$ diverges as $J/J'$ and, by using a simple algebra, one can show that

\begin{equation} \label{P1_Jimp}
  \lim_{\frac{J'}{J} \rightarrow 0}  p^{(1)}=1- \frac{4J'^2}{J^2} \sin^2(\frac{Jt}{2}).
\end{equation}

This clearly shows that the probability of success in measurement is always very close to 1, with an additional oscillatory term. At certain times, when $\sin(Jt/2)=0$, the probability $p^{(1)}$ is exactly 1. Another interesting fact is that the frequency of oscillations is determined by $J$, which allows for several oscillations within the transfer time $t_m \sim J'^2/J$.

This simple and exact analysis of a chain with four spins provides a good insight into the resonances that we observe in $P_{suc}$ for longer chains. Of course, for large chains the probability of success cannot be explained by a simple function like Eq.~(\ref{P1_Jimp}), but qualitatively the physics remains the same. The success probability of the protocol $P_{suc}$ can be optimized by a judicious choice of the measurement time $\tau$.

\section{Proposal for experimental realization}

One possible implementation of our proposed protocol could be realized with cold atom arrays in optical lattices. In such systems, two counter propagating laser beams create a regular potential with tuneable barriers, which can create a Mott insulator phase of atoms with exactly one particle per site \cite{Greiner-Mott}. In the limit of high on-site energy compared with the tunneling rate, the interaction between atoms is effectively modeled by a spin Hamiltonian \cite{Lukin-spin-Hamiltonian}.
Recently, local spin rotations and measurements \cite{Meschede-single-site,Bloch-single-site-2011} together with time resolved dynamics \cite{Bloch-single-site-2011} have been experimentally achieved. New advances in single site resolution in optical lattices \cite{Bloch-single-site-2011} made it possible to realize local rotations and measurements on individual atoms. Furthermore, the propagation of a single impurity spin \cite{Bloch-spin-wave} and magnon bound states \cite{Bloch-magnon} in a ferromagnetic spin chain have recently been experimentally realized. In spite of all these advances in optical lattices, there are still a few challenges which have to be overcome in order to realize our scheme, including local manipulation of the Hamiltonian and high fidelity atom detection. Nevertheless, the trend of the technology shows that these challenges could be overcome in a near future.

In order to realize our protocol, one has to first create the two impurities at the ends of a uniform spin chain. Initially the barriers between the atoms are high, i.e.\ there is no interaction, and all atoms are prepared in state $|0\ra$. Thanks to the local addressability of individual atoms with the single site resolution \cite{Bloch-single-site-2011}, this can be achieved by superimposing two extra potential wells, highly localized on the boundary sites, on the optical lattice potential \cite{Pachos-Knight-PRL-2003} as shown in Fig.~\ref{fig9}(a). Further rotations on the first atom can create an arbitrary state for the first qubit.
To realize such rotations, without affecting the neighboring qubits, one may apply a weak magnetic field gradient \cite{Meschede-single-site}, or use a focused laser beam \cite{Bloch-single-site-2011} to split the hyperfine levels of the target atom. Then, a microwave pulse --- tuned only for the target qubit --- operates the gate locally, as it has been realized in Refs.~\cite{Meschede-single-site,Bloch-single-site-2011}. When the initialization is accomplished, the interaction between the atoms is switched on through lowering the barriers and the dynamics starts. For spin measurements in the channel, one can use the technique in Ref.~\cite{Meschede-single-site}. According to that approach, state $|1\ra$ is coupled to an excited state through an intense perpendicular laser beam whose radiation pressure pushes the atom out of the lattice. This leaves the site empty if its atom is in state $|1\ra$, and full if the atom is in state $|0\ra$, as shown schematically in Fig.~\ref{fig9}(b). This can then be checked by global fluorescent imaging. The conclusion of the protocol, i.e.\ the final readout of the last qubit, is just another spin measurement.  It is worth mentioning that the measurement process is relatively slow, during which the system may evolve. In order to avoid this, one has to raise the barriers between the neighboring atoms to stop the dynamics and then perform the measurement. When the measurement is accomplished, one has to lower the barriers again to restore the dynamics.

\section{Concluding remarks}

We have proposed a protocol based on regular global measurements to protect a quantum spin channel, weakly coupled to the sender and receiver qubits, from the nocuous effects of dephasing. In fact, our measurement-assisted quantum communication protocol offers much higher transmission fidelities than the traditional no-measurement approach. This has the price that the process becomes probabilistic, but the probability of success can be maximized by tuning the rate of the measurements.

The success probability of our protocol oscillates with a frequency depending on the coupling of the channel Hamiltonian. And the quantum state transfer time is determined by the coupling of the effective Hamiltonian, which is much weaker than the channel couplings. This actually allows for several maxima in the success probability before the transfer is accomplished.

It is worth emphasizing that our proposed protocol works even for time-varying noise and for Markovian dephasing (modelled by a Lindbladian master equation), for both of which the well established dynamical decoupling approach fails. Furthermore, our proposal can also counter the effects of more complicated decoherence effects including those which create excitations in the channel through Lindblad operators like $\sigma^+$. Of course, these terms lower the probability of success in our mechanism as it is more likely to find an excitation in the channel.

\section*{Acknowledgements}

The authors thank Sougato Bose for his valuable comments. AB thanks the Physics of Information Group, at Instituto de Telecomunica\c{c}\~oes, in Lisbon, for the hospitality and acknowledges the support of the EPSRC grant $EP/K004077/1$. YO thanks the support from Funda\c{c}\~{a}o para a Ci\^{e}ncia e a Tecnologia (Portugal), namely through programmes PTDC/POPH and projects UID/Multi/00491/2013, UID/EEA/50008/2013, IT/QuSim and CRUP-CPU/CQVibes, partially funded by EU FEDER, and from the EU FP7 projects LANDAUER (GA 318287) and PAPETS (GA 323901).

\end{document}